\documentclass{article}
\usepackage{graphicx}
\usepackage[margin=2cm]{geometry} 
\usepackage{setspace}
\usepackage{amsthm,amsmath,amsfonts,amssymb} 
\usepackage{rotating,url} 
\usepackage{comment}
\usepackage{subfigure}
\usepackage{placeins}
\usepackage{xcolor}
\usepackage{authblk} 
\usepackage{fancyhdr} 
\usepackage{titlesec}

\usepackage[numbers,sort&compress]{natbib}

\title{Absolute Risk Prediction for Cannabis Use Disorder in Adolescence and Early Adulthood Using Bayesian Machine Learning}
\date{}
\newcommand{\keywords}[1]{\textbf{Keywords:} #1}

\author[1]{Tingfang Wang}
\author[2]{Joseph M. Boden}
\author[1,*]{Swati Biswas}
\author[1,*]{Pankaj K. Choudhary}
\affil[1]{Department of Mathematical Sciences, University of Texas at Dallas, Richardson, USA}
\affil[2]{Department of Psychological Medicine, University of Otago, Christchurch, New Zealand}
\affil[*]{Correspondence: Swati Biswas, 800 W Campbell Rd, FO 35, Richardson, TX 75025, USA. Email: swati.biswas@utdallas.edu OR Pankaj K. Choudhary, 800 W Campbell Rd, FO 35, Richardson, TX 75025, USA. Email: pankaj@utdallas.edu}

\begin{document}

\maketitle
\setstretch{1.5}

\begin{abstract}
\noindent\textbf{Introduction:} Substance use disorders (SUDs) have emerged as a pressing public health concern in the United States, with adolescent substance use often leading to SUDs in adulthood. Effective strategies are needed to stem this progression. To help fulfill this need, we developed a novel absolute risk prediction model for cannabis use disorder (CUD) for adolescents or young adults who use cannabis.

\noindent\textbf{Methods:} We trained a Bayesian machine learning model that provides a personalized CUD absolute risk for adolescents or young adults who use cannabis with data from the National Longitudinal Study of Adolescent to Adult Health. Model performance was assessed using 5-fold cross-validation (CV) with area under the curve (AUC) and ratio of the expected to observed number of cases (E/O). Independent validation of the final model was conducted using two datasets.

\noindent\textbf{Results:} The proposed model has five risk factors: biological sex, delinquency, and scores on personality traits of conscientiousness, neuroticism, and openness. For predicting CUD risk within five years of first cannabis use, AUC values for the training dataset and two validation datasets were 0.68, 0.64, and 0.75, respectively, and E/O values were 0.95, 0.98, and 1, respectively. This indicates good discrimination and calibration performance of the model.

\noindent\textbf{Discussion and Conclusion:} The proposed model can aid clinicians in assessing the risk of developing CUD among adolescents and young adults who use cannabis, enabling clinically appropriate interventions.

\end{abstract}


\newpage

\section{INTRODUCTION}
Substance use disorders (SUDs) are a growing public health concern worldwide. A report analyzing data from the Institute of Health Metrics and Evaluation database estimated the global prevalence of SUDs at 2.2\%, with alcohol use disorders being the most common at 1.5\% and drug use disorders at 0.8\%. Among drug use disorders, cannabis use disorder (CUD) has the highest prevalence, affecting 0.32\% of the global population, with the highest rates reported in Europe and the Americas \cite{globalsud}.

Cannabis is the most widely used illicit drug in both the United States (US) and Europe, particularly among adolescents and young adults \cite{globalsud}. In 2022, 22.0\% of individuals aged 12 or older (61.9 million) in the US reported cannabis use in the past year, with the highest prevalence (38.2\%) among young adults aged 18 to 25 \cite{SAMHSA2023}. In Europe, an estimated 15.0\% (15.1 million) of individuals aged 15 to 34 reported cannabis use in 2023 \cite{Europeansud}. The teenage years, in particular, are a critical window of vulnerability to substance use and SUDs \cite{NIDA2014}. As cannabis use in adolescence/youth is a precursor to the development of CUD later in life, there exists a window of opportunity to utilize the individual-level risk factors of adolescents or young adults who use cannabis to identify those at high risk of developing CUD and prevent CUD development. A potentially useful tool for this purpose is a statistical model that can take the profile of a person who uses cannabis and provide their personalized risk of developing CUD in a user-specified time interval.

Absolute risk prediction models are widely used in clinical settings for predicting the probability of an event of interest occurring within a specified time frame after adjusting for competing risks such as death from unrelated causes that preclude the event \cite{Gail2007, Pfeiffer2017}. Evaluation of the absolute risk of a disease or disorder plays a key role in developing health intervention strategies by identifying individuals who are at high risk \cite{gail2011, jackson2005, jackson2000}. At the individual level, absolute risk estimates can be used to counsel individuals based on their personalized risk profiles. Despite the widespread use of absolute risk prediction models across various diseases and disorders \cite{Gail2007, Chowdhury2017, dagostino2008, Freedman2009, Kovalchik2013, Pfeiffer2013, Sajal2022}, such models are not yet available for SUDs.

In fact, a recently proposed CUD risk prediction model by Rajapaksha et al. \cite{Rajapaksha2022}, to the best of our knowledge, is the only externally validated risk prediction model that was built using a nationally representative, longitudinal dataset. However, it is \textit{not} an absolute risk prediction model. It is a pure risk prediction model wherein competing causes of mortality were ignored. It provides the probability of developing CUD in the future for an individual who uses cannabis in an \textit{unspecified} time interval, limiting its practical utility for counseling. Since an individual is subject to mortality from competing risks, especially when the study period is long, a pure risk prediction model is less relevant for clinical decision-making \cite{Pfeiffer2017}. Another study provides the relative risk of CUD, which is the ratio of risk for an individual with certain levels of risk factors to risk for an individual with all risk factors set at baseline \cite{Richter2017}. While relative risks are useful for assessing the strength of risk factors, such models are not nearly as useful as absolute risk models for making clinical decisions or establishing policies for disease prevention \cite{Pfeiffer2017}. To address the growing concern of CUD in the US, particularly as the number of adolescents engaging in vaping and edible consumption of cannabis has increased over the years \cite{Patrick2020,lim2022}, an absolute risk prediction model is crucial for improving risk assessment and decision-making in both clinical and policy domains.

Numerous studies have identified a wide range of risk factors associated with SUDs, including CUD. These factors include male gender, early initiation of substances, exposure to traumatic events, peer substance use, engagement in delinquent acts, and personality traits such as conscientiousness and openness \cite{Agrawal2014, Cleveland2008, Connor2021, Kelly2019, Vanderpol2013, Verweij2010, roche2019, robinson2022}. Given the multitude of potential risk factors, it is essential that a risk prediction model be parsimonious for it to be amenable to adoption in clinical settings. Regularization can help in achieving this goal by shrinking the regression coefficients towards zero \cite{Tibshirani1996}. Bayesian methods offer a flexible framework by allowing regularization as well as estimation of model parameters jointly and accounting for uncertainty in their estimation via posterior distributions \cite{Tibshirani1996, Park2008, Van2019}.

In this study, we build a Bayesian machine learning model aimed at providing personalized estimates of the absolute risk of developing CUD for adolescents or young adults who use cannabis, based on their risk factor profiles. It considers a comprehensive set of risk factors from literature that are known to be associated with CUD. Additionally, to assess the model’s generalizability and robustness, we validate it on two independent datasets including an external dataset. 

\section{METHODS}

\subsection{Data sources}

We used data from two different sources for model training: National Longitudinal Study of Adolescent to Adult Health (Add Health) \cite{Harris2019} and the United States Life Tables, provided by the National Center for Health Statistics \cite{Arias2022united}. More specifically, we relied on the Add Health data to estimate the CUD hazard and the life table data to estimate the mortality hazard rate.

Add Health is a longitudinal study of a nationally representative sample of over 20,000 adolescents, consisting of rich demographic, social, familial, socioeconomic, behavioral, psychosocial, cognitive, and health survey data collected in five waves from 1994 to 2018 \cite{Harris2019}. Our training data consisted of participants who used cannabis, participated in wave IV (when questions related to CUD diagnosis were asked), reported the age of first use of cannabis and age of CUD onset (for those with CUD), reported same age of cannabis initiation in waves I and IV when this question was asked (for those who initiated before wave I), and had a wave I survey weight available. This process yielded a sample of 8,169 individuals with information such as the age of first use of cannabis, CUD status, age of CUD onset or censoring, ages at different waves, survey weights, and potential predictors. This sample served as our study cohort. The follow-up duration for each participant in this cohort commenced at the age of first use of cannabis. For participants who developed CUD, the follow-up concluded at the age of CUD. For others, the follow-up concluded at wave IV. Thus, age at wave IV is the censoring age for CUD-free individuals (wave V did not have any CUD related questions).

The United States life tables provided mortality patterns and life expectancy information within the US. For our study, we employed the most up-to-date publication available at the time of model training, from the year 2020 \cite{Arias2022united}.

We validated the proposed model using two independent datasets. The first validation dataset consisted of Add Health participants who met all inclusion criteria except for the availability of survey weights because of which they could not be included in the training dataset. The second dataset came from the Christchurch Health and Development Study (CHDS), which collected data on health, education, and life progress for a cohort of 1,265 children born in Christchurch, New Zealand, in 1977 \cite{Poulton2020}. As CHDS is an external dataset and the risk factors for CUDs are similar in New Zealand \cite{fergusson2015,fergusson2000}, validation on these data can demonstrate portability and robustness of the model.

\subsection{Absolute risk}
Let $T_1$ be the age of CUD onset and $T_2$ be the age of death from causes other than CUD. While there is no assumption of independence between $T_1$ and $T_2$, it is assumed that only one event can occur at a given time \cite{Pfeiffer2017}. Let $\delta(t)$ be a state process taking values in $\{0,1,2\}$ that describes the state of a person at time $t$. In particular, an individual who uses cannabis and is free of CUD is in state $0$ until one of the two events occurs. Thus, the observed time $T=\min\{t>0, \delta(t)\neq0\}$ is the first time an individual is not at the initial state and $\delta(t)$ indicates the type of the event that occurred at time $T$. We denote the cause-specific hazard function as $\lambda_m(t)=\lim_{\epsilon\to0^+}\frac{P(t\leq T<t+\epsilon, \;\delta(T)=m|T\geq t)}{\epsilon}$, $m = 1,\;2$. Under the assumption that only one event can occur at a given time, the overall hazard function of $T$ is $\lambda(t)=\lambda_1(t)+\lambda_2(t)$. For an individual who uses cannabis and is free of CUD at age $a$, the absolute risk of getting CUD by age $b$ ($b>a$) is given by \cite{Pfeiffer2017}:
\begin{align}\label{absrisk}
    r(a,b)&=P(a<T\leq b,\; \delta(T)=1|T\geq a)\nonumber\\
    &=\int_a^b \lambda_1(t) \exp\left\{-\int_a^t[\lambda_1(u)+\lambda_2(u)]du\right\}dt.
\end{align}

We note that the competing risk of death due to non-CUD causes is small in our specific target population, but we incorporated it for methodological completeness and generalizability to other domains.

\subsection{Modeling framework}
 
Assume that the CUD hazard rate $\lambda_1$ follows a Cox proportional hazard (PH) model \cite{Cox1972} given a set of risk factors, some of which are measured in a single wave (cross-sectional) while others are measured in multiple waves (longitudinal). For each longitudinal predictor, we first fit a linear mixed model where the predictor is modeled as a function of age. The estimated random intercept from this model is then used as a predictor in the CUD risk prediction model \cite{chen2015}. More details can be found in Supplement Section S1. In addition to using random intercept, later we also explore a simplified version by using average of the longitudinal predictor values over waves waves as that will be much simpler to implement in clinical settings \cite{chen2015}. We note that while time-varying covariates are commonly used in survival analysis models, extending such models for absolute risk prediction is not trivial and will require additional methodological development, which is beyond the scope of this article. Moreover, in our training data, longitudinal predictors were measured at most three times with substantial amount of missingness, making their direct inclusion difficult from practical point of view. The baseline hazard was modeled using M-splines to ensure monotonicity of the corresponding cumulative baseline hazard \cite{Ramsay1988}. The mortality hazard rate $\lambda_2(t)$ from non-CUD causes was estimated by all-cause mortality hazard rate under the assumption that death due to CUD is rare \cite{UnderlyingDeath, NAS2017}. Details about the modeling framework are provided in Supplement Section S2.

\subsection{Estimation framework}
The task of estimating the absolute risk expressed in equation \eqref{absrisk} reduces to that of estimating the CUD hazard function $\lambda_1(t)$. We employed a Bayesian machine learning approach within the Cox PH model framework to estimate this hazard function. Specifically, we used a lasso prior to regularize the regression coefficients. For the baseline hazard M-spline coefficients, we used a Dirichlet prior. To account for the complex survey design, we incorporated each participant’s survey weight into the likelihood function. The joint posterior distribution was estimated using a Markov chain Monte Carlo algorithm. The fitted model can be used for predicting the absolute risk of developing CUD in a certain time frame for an individual who uses cannabis and is free of CUD. Supplement Sections S3 and S4 provide details about these steps in estimation and prediction.

\subsection{Model building steps}

From the Add Health questionnaires, we derived 45 potential risk factors for CUD. Some of them had a substantial amount of missing values. To identify the most influential risk factors while maximizing the sample size for fitting multiple regression models, we employed the following four-step approach:

(1)	Initial Screening: We fitted univariate Cox PH model for each potential risk factor and selected those with p-values less than 0.25.

(2)	Multivariate Screening: Subsequently, we fitted multivariate Cox PH models with various combinations of the risk factors selected in step (1). The combinations were chosen to ensure that at least 80\% of the initial cohort had complete data on the selected risk factors. The predictors with p-values below 0.25 in at least one of those fitted models were retained for further consideration \cite{hosmer2000}. 

(3) Bayesian Model Fitting: We fitted the Bayesian model as described in “Estimation Framework” using the risk factors identified in step (2). For variable selection, we employed two methods: credible interval and scaled neighborhood \cite{VarSelect2010}. This step resulted in several unique competing models obtained by varying confidence levels and thresholds for the two methods.

(4)	Model Comparison: Lastly, we compared the performance of the competing models identified in step (3) using 5-fold cross-validation (CV) to identify the optimal absolute risk prediction model. 

Additional details regarding these steps are provided in Supplement Section S5. The model comparison was based on two primary metrics: the area under the receiver operating characteristic curve (AUC) and the ratio of expected (E) to observed (O) number of events (E/O) \cite{Pfeiffer2017}. Two types of prediction were considered for this assessment: (1) risk of CUD within a specified number of years following the initiation of cannabis use, and (2) risk of CUD within a certain number of years for participants of a particular age, if they had started using cannabis by that age. 

\section{RESULTS}

\subsection{Fitted models}

Our study cohort, as detailed in Data sources, comprised 8,169 participants, with lifetime CUD prevalence of 8.2\%. In the model-building step (2), 21 risk factors were identified (details on these risk factors can be found in Supplement Table S1). However, due to missing data in several variables, the training sample size for the optimal model found in step (4) was reduced to 8,068, with a slightly lower lifetime CUD prevalence of 8.18\%.

A total of six competing models were identified in step (3). These are listed in Supplement Table S2. The optimal model among them as identified by 5-fold CV has five risk factors: biological sex, delinquency, and scores on three personality traits — conscientiousness, neuroticism, and openness. As delinquency is a longitudinal predictor, for ease of implementation in clinical settings, we investigated the possibility of replacing the estimated random intercept by the average of delinquency values across the waves \cite{chen2015}. We found that this replacement, in fact, improved the model performance in terms of AUC and E/O. Moreover, using the mean value simplifies both the implementation and interpretation of the model. The results of this model with delinquency scale measured as mean over waves are presented in Table \ref{t:t1}. Males, higher levels of neuroticism, openness, and delinquency scale, along with lower conscientiousness scale, were found to be associated with higher risk of CUD. We refer to it as Model 1. 

The risk factor welfare status dropped out in the model building step (3). As the influence of socioeconomic status on CUD is widely documented in the literature and is of interest in both scientific research and public health \cite{Calling2019, Goodman2002, Karriker2011, Karriker2013, Lee2015}, we explored a variant of Model 1 by incorporating welfare status as an additional risk factor. The missing data in the welfare variable reduced the training sample size to 7,954 with lifetime CUD prevalence of 7.3\%. Table \ref{t:t1} shows the results of this model, denoted by Model 2, indicating that higher risk of CUD is associated with receipt of welfare, in addition to the associations found in Model 1. We checked the proportional hazards assumption of both models using the Schoenfeld residual test and found that the assumption was reasonable, as judged by the test p-values.

\subsection{Model evaluation using 5-fold CV}

The predictive performance of Models 1 and 2 is summarized in Supplement Table S3. For Model 1, AUC ranged from 0.62 to 0.72 across various age intervals, with the highest value achieved in 1-year predictions made at the age of first cannabis use and 1-year predictions made at age 18. Approximately 82\% of E/O values were between 0.8 to 1.2. Similarly, for Model 2, AUC ranged from 0.65 to 0.72 across different age intervals for predictions, with the highest values observed for the same predictions as for Model 1. Moreover, about 85\% of the E/O values were between 0.8 to 1.2. These results indicate good discrimination and calibration abilities of the models.
 
\begin{figure}[ht]
\begin{center}
\includegraphics[scale=0.6] {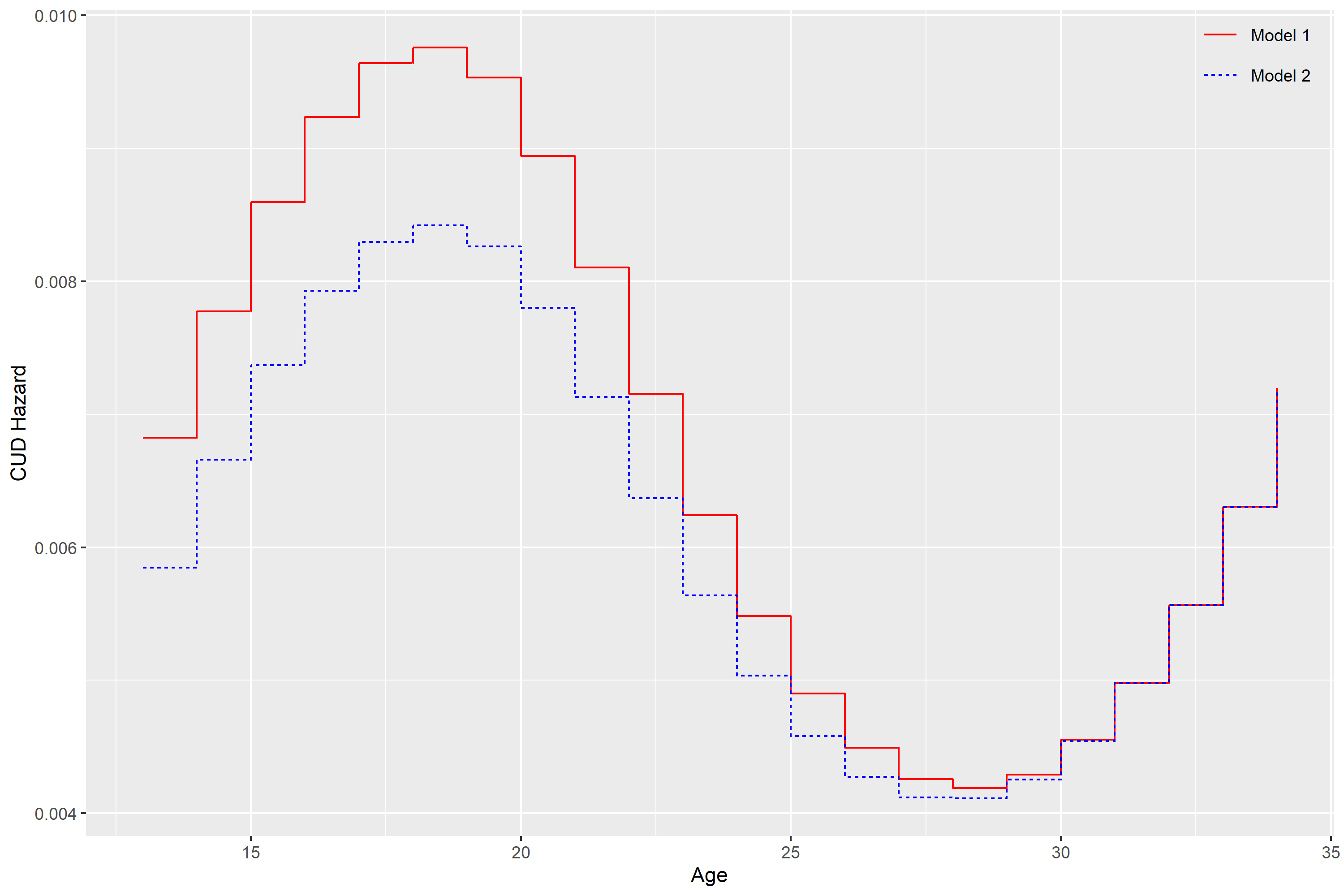}
\caption{CUD hazards.}
\label{fig:f1}
\end{center}
\end{figure}

In Figure \ref{fig:f1}, the estimated average CUD hazard is shown across ages 13 to 34 for both models. They exhibit similar patterns, initially increasing from age 13 to 18, followed by a decline until approximately age 27, before a subsequent increase. This pattern is consistent with research indicating that adolescents are more susceptible to SUDs \cite{NIDA2014}, with the highest percentage of CUD observed among young adults aged 18 to 25, followed by those aged 12 to 17 \cite{SAMHSA2023}. The hazard estimate under Model 1 is higher than that under Model 2 until about age 30, after which the two are identical. Their difference is maximum between the ages of 17 and 19.

\section{INDEPENDENT VALIDATION}

We validated the proposed models using two independent test datasets described in Section 2.1: (1) Add Health test data and (2) CHDS data. For CHDS data, although participants were followed up to 40 years of age with assessments at various ages, we only used their data up to age approximately 30 to maintain consistency with the training data \cite{Poulton2020, Lakishka2023}. For each individual in these datasets, the prediction was done in the same way as described in Section 2. Then AUC and E/O were computed for evaluating model performance.

\subsection{Validation on Add Health test data}

\subsubsection{Sample characteristics}

The data used for validating Model 1 comprised 520 participants, with lifetime CUD prevalence of 7.9\%, similar to the prevalence of 8.18\% in the training data for Model 1. For the validation of Model 2, the sample size was reduced to 514 due to missing values in welfare, with associated lifetime CUD prevalence of 7\%, which is comparable to the prevalence of 7.3\% in the Model 2 training data.

\subsubsection{Validation results}

\begin{figure}[ht]
\begin{center}
\includegraphics[scale=0.6] {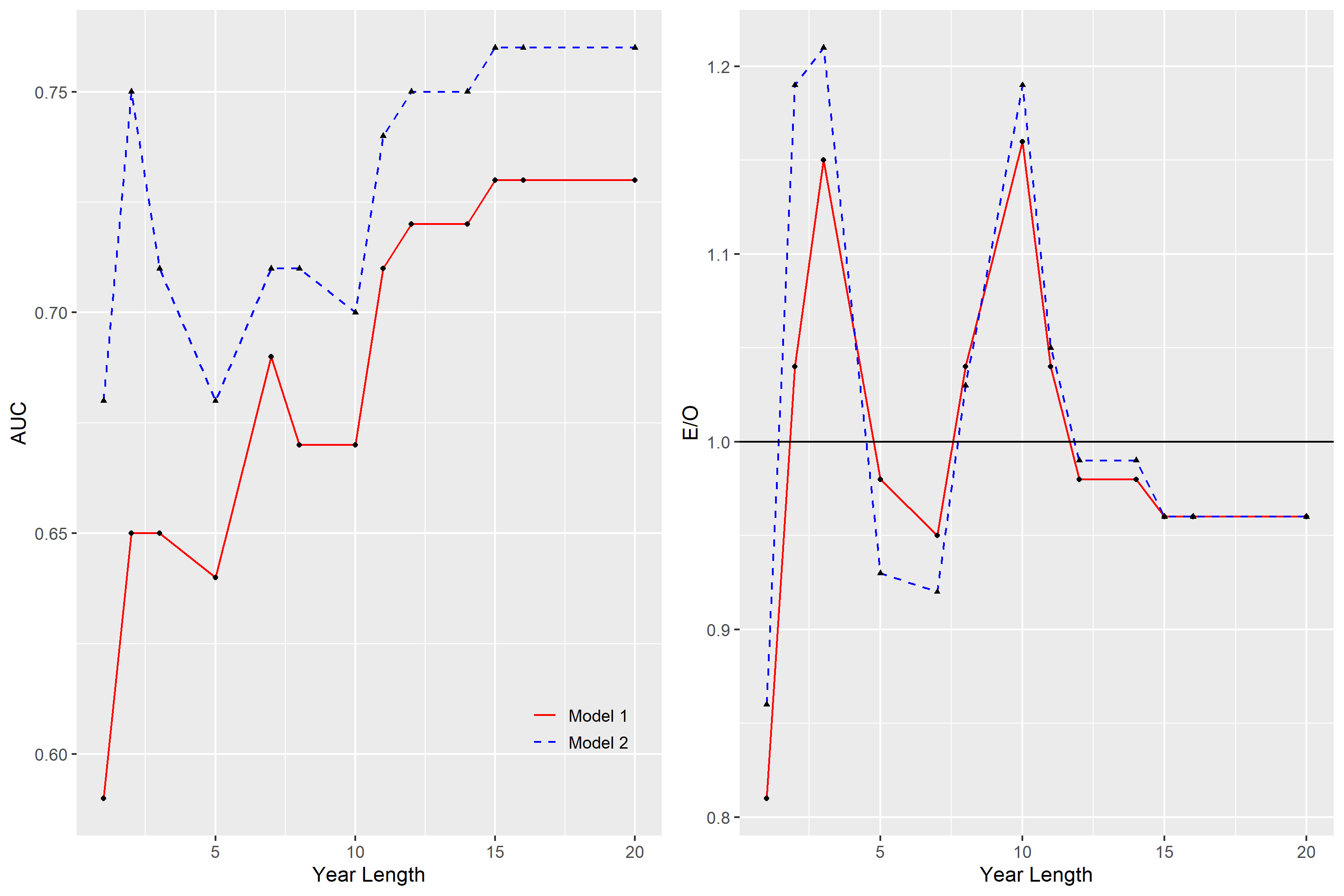}
\caption{Validation on Add Health test data: AUC and E/O trajectories for CUD risk prediction within a certain number of years made at the age of first cannabis use.}
\label{fig:f2}
\end{center}
\end{figure}

Figure \ref{fig:f2} illustrates the trajectories of AUC and E/O for predicted risk of CUD within a certain number of years of the age of first cannabis use for both models. Model 2 generally exhibited higher AUC values, particularly for shorter prediction intervals after the initial cannabis use. However, as the prediction duration exceeded 5 years, the differences between the models became small. The E/O values of the two models were comparable with no values deviating substantially from 1. Supplement Figure S1 depicts the same trajectories for predictions within a certain number of years of specific ages. Here, both models showed similar AUC and E/O values, except AUC values for 1-year predictions made at age 15. In this case, the number of CUD cases is small.

Supplement Table S4 presents E/O values for 5-year predictions made at the age of first cannabis use across four risk quartile groups and different levels of predictors. The continuous predictors were categorized using their median values. For Model 1, E/O values were between 0.8 to 1.2 for the third and fourth risk quartile groups, both biological sexes, both levels of conscientiousness and openness, and above median delinquency group. A larger deviation from 1 was observed for the above median neuroticism group. For Model 2, E/O values remained between 0.8 to 1.2 for the third and fourth risk quartile groups, both biological sexes, below median conscientiousness, and both welfare groups. For the remaining groups, the values did not deviate too far from 1 except for the second risk quartile group.

Supplement Table S5 shows E/O values for 5-year predictions made at age 16 for the same risk quartile groups and predictor levels. For Model 1, E/O values were between 0.8 to 1.2 for all risk quartile groups except the third quartile, both biological sexes, below median conscientiousness, above median delinquency, both levels of neuroticism and openness. No large deviations were observed for the remaining groups. For Model 2, E/O values remained between 0.8 to 1.2 for the fourth risk quartile group, females, below median conscientiousness and openness, above median neuroticism and delinquency, and both welfare groups. Similar to Model 1, no large deviations from 1 were observed for the other groups. On the whole, these findings indicate that both models generally perform well on Add Health test data in terms of model discrimination and calibration.  

\subsection{External validation on CHDS data}

\subsubsection{Sample characteristics}

A dataset consisting of 637 individuals who use cannabis was previously used by Ruberu et al. \cite{Lakishka2023} to validate a Bayesian logistic regression model for predicting CUD risk \cite{Rajapaksha2022, Lakishka2023}. We used the same dataset for validating our proposed absolute risk prediction model. A more detailed description of the dataset and construction of the predictors can be found in \cite{Lakishka2023}. However, due to unavailability of welfare information in this dataset, we could only validate Model 1. The lifetime prevalence of CUD in these data was 16.5\%, which is notably higher than the prevalence of 8.18\% observed in the training dataset.

Given this discrepancy and following Ruberu et al. \cite{Lakishka2023}, we explored the possibility of model recalibration as suggested in literature \cite{Pfeiffer2017, recalibration2, recalibration1}. Specifically, we applied logistic re-calibration to update the intercept in the model for log odds of CUD risk to reflect the high prevalence in the CHDS validation data. Note that even though this re-calibration method leads to updated absolute risks, the relative rankings of those individual risks do not change, thereby leaving the AUC unaffected. Details about recalibration can be found in Supplement Section S6.

\subsubsection{Validation results}

\begin{figure}[ht]
\begin{center}
\includegraphics[scale=0.5] {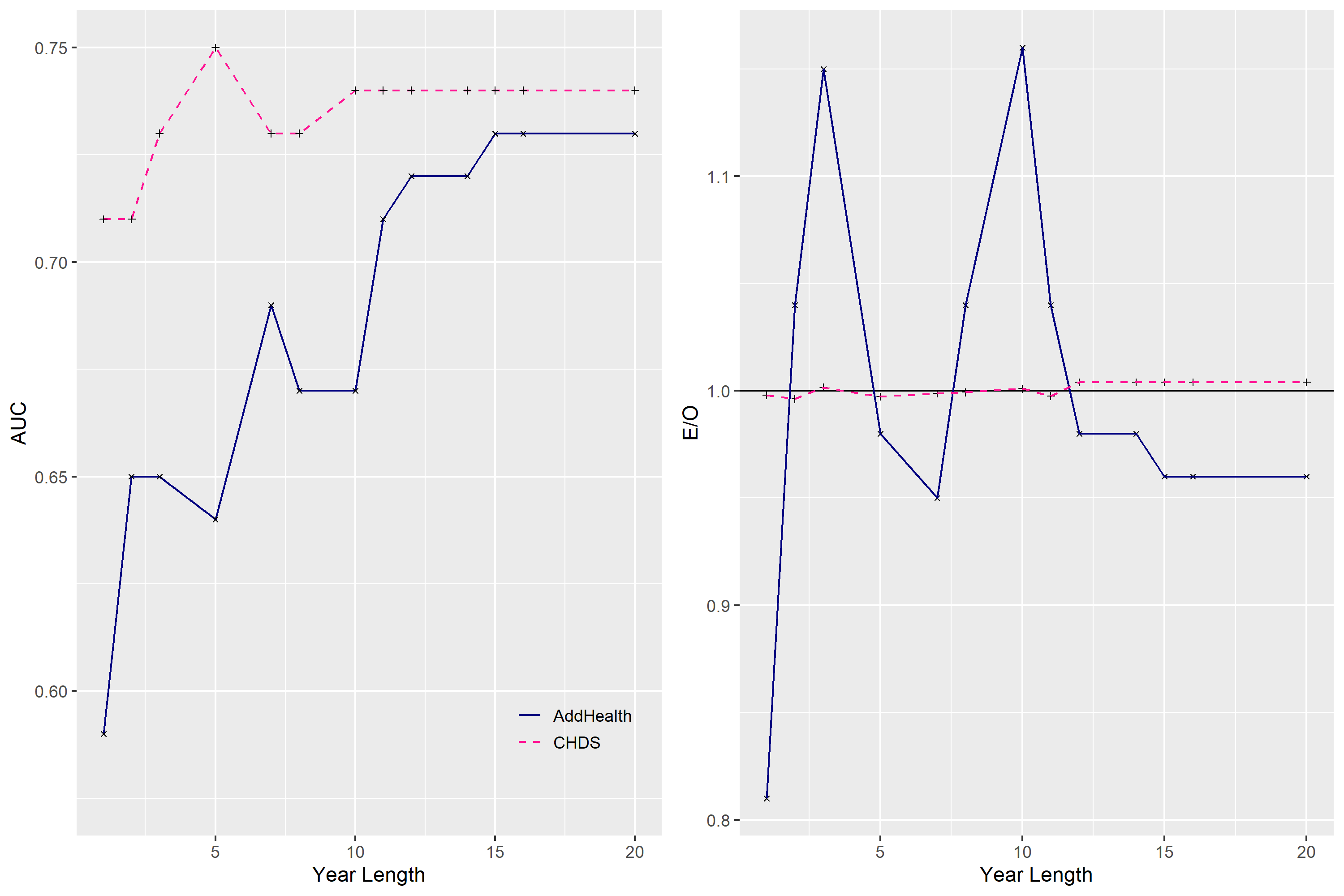}
\caption{AUC and E/O trajectories of Model 1 for CUD risk prediction within a certain number of years made at the age of first cannabis use for both validation datasets.}
\label{fig:f3}
\end{center}
\end{figure}

Figure \ref{fig:f3} shows the AUC and E/O trajectories for predicted risk of CUD within a certain number of years of the age of first cannabis use. Supplement Figure S2 shows the same trajectories for predictions within a certain number of years made at specific ages (the results from Add Health test data are also plotted for ease of comparison). The AUC values fell between 0.71 to 0.75 in predictions within a certain number of years made at the age of first cannabis use. As for the other type of prediction, most AUC values fell between 0.65 to 0.75 (the results are also shown in Supplement Table S7). All E/O values were close to 1 (for CHDS) after recalibration. These results indicate good predictive performance of the model.

Comparing the results from the CHDS data with that from the Add Health test data, we see that in predictions within a certain number of years of the age of first cannabis use, AUC values for CHDS were consistently higher. However, the difference tended to decrease with longer prediction intervals. For predictions within a certain number of years of specific ages, the AUC values of both datasets were comparable, especially for longer prediction intervals. As noted above, E/O values for the CHDS data were close to 1, while for Add Health, the values did not substantially deviate from 1, particularly for prediction intervals longer than 2 years.

Supplement Table S6 provides E/O values for 5-year predictions made at the age of first cannabis use and at age 16 across the four risk quartile groups and different levels of each predictor. For predictions made at the time of cannabis initiation, E/O values ranged from 0.8 to 1.2 for the third risk quartile group, above-median delinquency and openness, all conscientiousness, and all neuroticism groups. Larger deviations were observed for the first risk quartile group and females. For predictions made at age 16, E/O values were between 0.8 to 1.2 for all quartile groups except the third quartile, below median conscientiousness and neuroticism, both levels of openness, and above median delinquency groups. E/O values deviated more from 1 for females compared to males. Altogether, these findings indicate that Model 1 has generally good discrimination and calibration performance on CHDS data. 

\section{DISCUSSION}

Cannabis is the most widely used illicit drug, especially among adolescents and young adults \cite{engelgardt2023, hammond2020}. With growing legalization of non-medical cannabis use, CUD prevalence may increase \cite{NAS2017, berenson2019, ncsl2024cannabis}. A CUD risk prediction model can aid clinicians in identifying high risk individuals who can then be helped with early and age-appropriate interventions. Absolute risk prediction models are most clinically relevant for this task. Yet such models are non-existent for CUD.

To address this gap, we developed two absolute risk models that provide individualized risk of CUD. Model 1 consists of just 5 risk factors, whereas Model 2 additionally has welfare status. This risk factor was added to Model 1 because the relationship of SUD with social inequality is of current scientific interest \cite{Karriker2011, amaro2021, baptiste2018}. Since both models perform equally well, based on model parsimony, we consider Model 1 as our final proposed model. Model 2 can be also used when welfare status is available. Our findings are in general alignment with research literature. Specifically, CUD tends to be more common in males \cite{Hayatbakhsh2009multiple, Meier2016which}, those experiencing socio-economic stress and deprivation \cite{Room2005, WOHLFARTH199851}, individuals with higher levels of neuroticism and lower levels of conscientiousness trait, and those coming into contact with the police and justice systems at earlier ages \cite{RAJAPAKSHA2020, Zoboroski2021classical, Koh2017violence}. The robust performance of the models in terms of both discrimination and calibration in independent data including an external dataset validates their practical utility and potential for informing effective public health interventions, thereby improving outcomes for individuals who use cannabis.

In comparison to the logistic regression model of Rajapaksha et al. \cite{Rajapaksha2022}, our final proposed model is more parsimonious. Besides the five risk factors of our model, the logistic model additionally has two more: adverse childhood experiences and peer cannabis use. Yet, even with a smaller number of risk factors, our model has similar AUC values in independent validation on both Add Health and CHDS datasets \cite{Rajapaksha2022, Lakishka2023}. More importantly, our model provides absolute risk estimates of CUD for individuals within specific age intervals. While the logistic model simply predicts the risk of CUD in future without regard to any time interval and without considering the competing risk of death, making it less relevant for clinical decision-making. We computed 5- and 15-year absolute risks of CUD made at the age of first cannabis use for a diverse range of risk factor combinations and compared them with the corresponding probability of developing CUD in (unspecified) future as predicted by the logistic model \cite{Rajapaksha2022} in Table \ref{t:t3}. We can see that the absolute risks show considerable variability depending on the combination of risk factors and the length of the prediction period, which could not be captured by the logistic model of Rajapaksha et al. \cite{Rajapaksha2022}.

The parsimonious feature of our model makes it easier to implement it in clinical settings, especially for the sensitive target population. Another helpful feature with clinical relevance is that for a longitudinal predictor, our model requires just average across the available time points, making it applicable even in settings with just one single time point. Moreover, the model can be re-calibrated for other populations, if needed, as we did for CHDS data. For this, the only information needed is CUD prevalence in the new population, which is typically available from published governmental reports.

Our study has one limitation: the Add Health data are relatively dated, potentially limiting its ability to capture current trends in cannabis use, particularly in the light of recent legalization of non-medical cannabis use in many parts of the US \cite{berenson2019}. However, there is currently no other nationally representative longitudinal study available that has adolescent-to-adult follow-up as well as SUD outcomes and an extensive range of potential risk factors.

Despite this limitation, we believe the proposed model can be helpful in identifying adolescents and young adults at high risk of CUD, who can then be helped early on. The individualized absolute risk estimates can enhance clinician-patient interactions, improving patient education and decision-making. The model helps personalize treatment and prevention strategies, ensuring more targeted care and better resource allocation, ultimately improving patient outcomes and contributing to public health efforts to reduce CUD. Moreover, the methodology we proposed can be adapted to develop absolute risk prediction models for other diseases and disorders. There are, undoubtedly, ethical considerations for the sensitive age groups of adolescents that must be made before implementing this type of model in clinical settings \cite{lawrie2019ethical, leadbeater2018ethical}. This will require careful discussion with the clinicians who directly work with this target population. 

\section{CONCLUSION}

This study developed a novel, and to the best of our knowledge, the first-ever absolute risk model for predicting the risk of CUD in adolescents or young adults who use cannabis, based on personalized risk factors. A higher risk of CUD is associated with males (biological sex), higher levels of neuroticism, openness, and delinquency, along with lower conscientiousness. By providing individualized risk estimates within specific age intervals, our model offers a clinically relevant tool that can enhance early detection and intervention efforts, helping adolescents or young adults who use cannabis avoid the path to CUD.

\section{ACKNOWLEDGMENT}

The data used in this work are from Add Health and CHDS. Add Health is funded by grant P01-HD31921 from the Eunice Kennedy Shriver National Institute of Child Health and Human Development, with cooperative funding from 23 other federal agencies and foundations. CHDS is supported by the Health Research Council of New Zealand (Grant 16-600: The Christchurch Health and Development Study: Birth to 40 Years). This analysis did not receive direct support from grant P01-HD31921 or 16-600. The authors thank Dhanushka Rajapaksha for helping with data processing.  

\section{CONFLICTS of INTEREST} 

The authors have no interest to declare.

\bibliographystyle{unsrtnat} 
\bibliography{ref}

\newpage
\begin{table}[ht]
  \caption{Posterior means of hazard ratio (HR) and 95\% credible interval (CrI) for HR.}
   \label{t:t1}
  \begin{center}
  \begin{tabular}{lccp{0.1cm}cc} \hline
  \textbf{} & \multicolumn{2}{c}{\textbf{Model 1}} &  & \multicolumn{2}{c}{\textbf{Model 2}} \\ 
  \cline{2-3} \cline{5-6}
  \textbf{Variable} & \textbf{HR} & \textbf{CrI} &  & \textbf{HR} & \textbf{CrI} \\
   \hline
Biological sex & 1.31  & (1.12, 1.55)  & &1.27  &  (1.07, 1.51)  \\
Conscientiousness scale & 0.34  & (0.19, 0.58)  & &0.34  &(0.19, 0.61)  \\
Neuroticism scale &5.64 &(3.25, 9.68)  &  &6.11& (3.46, 10.91)  \\
Openness scale & 5.16 &(2.72, 9.77)   & & 5.99 & (2.97, 11.82)   \\
Delinquency scale & 19.89 & (12.06, 32.79) && 18.73 & (10.80, 32.14) \\
Welfare & &   & &1.03   &(0.85, 1.23)   \\
\hline
\end{tabular}
\end{center}
\end{table}

\begin{sidewaystable}[ht]
    \centering
    \caption{Predictions of 5- and 15-year absolute risk of CUD made at the age of first cannabis use and probability of CUD in future obtained from the logistic model\textsuperscript{1} for users with different risk profiles}
    \vspace{10pt} 
    \label{t:t3}
    \begin{tabular}{ccccccccccccc} \hline
        \textbf{Biological} & \textbf{Delinquency} & \textbf{Neuroticism} & \textbf{Openness} & \textbf{Conscient-} & \textbf{ACE} & \textbf{Peer} & \textbf{Age} & \textbf{CUD} & \textbf{CUD/} & \textbf{5-year} & \textbf{15-year} & \textbf{Probability\textsuperscript{1}}\\ 

        \textbf{sex} & \textbf{scale} & \textbf{scale} & \textbf{scale} & \textbf{iousness} & \textbf{scale\textsuperscript{3}} & \textbf{cannabis} & \textbf{first} & \textbf{status} & \textbf{wave IV} & \textbf{risk(\%)} & \textbf{risk(\%)} & \textbf{(\%)}\\
        
        \textbf{} & \textbf{} & \textbf{} & \textbf{} & \textbf{scale} & \textbf{} & \textbf{use\textsuperscript{3}} & \textbf{use} & \textbf{} & \textbf{age\textsuperscript{2}} & \textbf{} & \textbf{} & \textbf{}\\ \hline

Female & 0.02 & 0.35 & 0.45 & 1.00 & 0.11 & 0.00 & 18 & No & 26 & 0.94 & 1.85 & 17.04\\
Male & 0.06 & 0.25 & 0.50 & 0.90 & 0.00 & 0.33 & 16 & No & 30 & 1.51 & 3.06 & 28.93\\
Male & 0.93 & 0.60 & 0.65 & 0.80 & 0.11 & 0.33 & 16 & Yes & 17 & 43.89 & 69.73 & 96.98\\
Male & 0.56 & 0.75 & 0.85 & 0.55 & 0.44 & 1.00 & 14 & Yes & 20 & 34.02 & 62.04 & 97.13\\
Female & 0.06 & 0.55 & 0.70 & 0.70 & 0.00 & 0.50 & 15 & No & 28 & 3.27 & 6.89 & 49.64\\
Male & 0.06 & 0.55 & 0.70 & 0.70 & 0.00 & 0.50 & 15 & No & 28 & 4.27 & 8.93 & 57.58\\
Male & 0.03 & 0.55 & 0.60 & 0.65 & 0.11 & 0.17 & 15 & No & 29 & 3.51 & 7.38 & 47.37\\
Male & 0.20 & 0.55 & 0.60 & 0.65 & 0.11 & 0.17 & 15 & Yes & 16 & 5.79 & 12.00 & 63.40\\
Female & 0.05 & 0.50 & 0.80 & 0.80 & 0.33 & 0.00 & 18 & No & 26 & 2.87 & 5.63 & 42.59\\
Female & 0.05 & 0.60 & 0.80 & 0.80 & 0.33 & 0.00 & 18 & Yes & 24 & 3.41 & 6.67 & 46.87\\
Male & 0.01 & 0.40 & 0.70 & 0.80 & 0.22 & 0.00 & 18 & No & 29 & 2.39 & 4.70 & 37.32\\
Male & 0.01 & 0.40 & 0.95 & 0.80 & 0.22 & 0.00 & 18 & Yes & 19 & 3.58 & 7.00 & 47.60\\
Female & 0.07 & 0.40 & 0.65 & 0.80 & 0.11 & 0.00 & 16 & No & 29 & 2.18 & 4.39 & 32.39\\
Female & 0.07 & 0.40 & 0.65 & 0.35 & 0.11 & 0.00 & 16 & Yes & 27 & 3.56 & 7.11 & 44.56\\ \hline
    \end{tabular}
    \footnotetext{\textsuperscript{1} From Rajapaksha et al. (2022) \cite{Rajapaksha2022}.}
    \footnotetext{\textsuperscript{2} For CUD-free individuals (status = No), this is the age at wave IV (last follow-up).}
    \footnotetext{\textsuperscript{3} This risk factor is used in the logistic model only.}
\end{sidewaystable}

\FloatBarrier

\end{document}


\maketitle
\setstretch{1.5}

\titleformat{\section}[block]{\normalfont\bfseries}{S\thesection}{1em}{}
\titleformat{\subsection}[block]{\normalfont\bfseries}{S\thesection.\thesubsection}{1em}{}
\titleformat{\subsubsection}[block]{\normalfont\bfseries}{S\thesection.\thesubsection.\thesubsubsection}{1em}{}

\renewcommand{\thesection}{\arabic{section}}
\renewcommand{\thesubsection}{\arabic{subsection}}
\renewcommand{\thesubsubsection}{\arabic{subsubsection}}

\section{Summarizing Longitudinal Predictors by Participant-specific Random Intercept}
Let $l$ be a longitudinal predictor with $l_{it}$ denoting its value for participant $i$ in wave $t$. Here $i=1,\dots,n$ and $t \in \{1,2,3\}$, but some predictors were measured in two waves only so for such predictors the modeling is based on observations in two waves only. The predictor is modeled as a function of the participant’s age using a linear mixed model, formulated as:

\[
l_{it} = \gamma_0 + \nu_{0i} + \gamma_1 \text{age}_{it} + \epsilon_{it},
\]

\noindent where $\text{age}_{it}$ is the age of the $i^{th}$ participant at wave $t$, $\gamma_0$ and $\gamma_1$ are fixed effects, $\nu_{0i}$ is the random intercept of the predictor for the $i^{th}$ participant, and $\epsilon_{it}$ is the random error term. The random intercept is assumed to follow a normal distribution $N(0,\sigma_{\nu_0}^2)$. The random error is assumed to follow $N(0,\sigma_\epsilon^2)$ distribution and is independent of the random intercept. The estimated best linear unbiased predictor (BLUP) of the random intercept $\nu_{0i}$ is then incorporated as a covariate in the proposed CUD absolute risk prediction model. The lme4 package in R is used for model fitting \cite{lme42015}. 

\section{Details of the Modeling Framework}
\subsection{CUD Hazard Rate} 
Assume that the CUD hazard rate $\lambda_1$ follows a Cox proportional hazard (PH) model given a set of risk factors, $X$, and their effects denoted by $\beta$ \cite{Cox1972}:

$$ \lambda_1(t|X) = \lambda_0(t) \exp(X^\top \beta), $$

\noindent where $\lambda_0(t)$ is the baseline hazard function, that is, hazard of an individual at time $t$ with all risk factors set at baseline or zero, and $\exp(X^\top \beta)$ denotes the relative risk of an individual. 

We model the baseline hazard rate using M-splines \cite{Ramsay1988} to capture complex variations in the hazard rate over time while accommodating potential non-linearity. The M-spline function of degree $\eta$ with $L$ basis terms and $J$ knot locations $\kappa = \{\kappa_1, ..., \kappa_J\}$ is defined as:

$$M(t; \gamma, \kappa, \eta) = \sum_{l=1}^{L} \gamma_l M_l(t;\kappa, \eta),$$

\noindent where $\gamma$ is a vector consisting of M-spline coefficients $\gamma_l$, $l = 1,...,L$. Subsequently, the CUD hazard function can be expressed as:

\begin{equation}\label{cudhazard}
    \lambda_1(t|X) = M(t; \gamma, \kappa, \eta) \exp(X^\top \beta)
\end{equation}

\noindent and the corresponding survival function of CUD \cite{Cox1972, SurvBook}:

\begin{equation}\label{cudsurv}
    S_1(t|X) = \exp\left(-I(t; \gamma, \kappa, \eta) \exp(X^\top \beta)\right),
\end{equation}

\noindent where $I(t; \gamma, \kappa, \eta) = \sum_{l=1}^{L} \gamma_l I_l(t;\kappa, \eta)$ is the corresponding integrated spline (I-spline) function with 

$$I_l(t; \kappa, \eta) = \int_{\kappa_1}^{t} M_l(u; \kappa, \eta) du.$$

\subsection{Mortality Hazard Rate}

Assume that the risk of death from causes other than CUD does not depend on the risk factors $X$ included in the above hazard model. Death due to CUD is, in fact, rare \cite{NAS2017, UnderlyingDeath} and so the mortality hazard rate in the population from non-CUD causes is similar to the one from all causes. For the latter, a population level estimate is easily available from the National Vital Statistics Reports \cite{Arias2022united}. Let $\lambda_2(t)$ be a piecewise constant all-cause mortal hazard rate obtained from the life table. As this rate is calculated using data from the entire US population, it is thus assumed to be a known quantity without error \cite{Pfeiffer2017}. Therefore, the all-cause mortal survival rate can be expressed as:

$$ S_2(t) = \exp\left(-\int_0^t \lambda_2(u) du\right).$$

Finally, given the risk factors, $X$, the absolute risk defined in equation (1) of the main text can be written as:
\begin{align}\label{absrisk2}
    r(a,b|X) &= \int_a^b \lambda_1(t|X) \exp\left(-\int_a^t [\lambda_1(u|X) + \lambda_2(u)] du\right) dt \nonumber \\ 
    &= \int_a^b \lambda_1(t|X) \frac{S_1(t|X)S_2(t)}{S_1(a|X)S_2(a)} dt.
\end{align}

\section{Details of the Estimation Framework}
\subsection{Data Likelihood}
The task of estimating the absolute risk expressed in equation \eqref{absrisk2} reduces to that of estimating the CUD hazard function $\lambda_1(t)$. The CUD event time is subject to not only right censoring from the loss of follow-up but also left truncation at the age of initiation of cannabis use, as individuals can be at risk of CUD only after they start using cannabis. Let $C$ be the censoring time, assumed to be independent of $T_1$ (age of CUD onset), and $T_0$ be the age at first use of cannabis. Then the observed event time is the minimum of the censoring time and the event time, $T = \min\{C, T_1\}$, $T \geq T_0$.

Let $D_j$ denote the observed data for $j^{th}$ individual, $j=1,\ldots, N$. Let $T_{1j}$, $T_{2j}$, and $T_{0j}$ be the age of CUD onset, age of death from causes other than CUD, and age of initiation of cannabis use, respectively. Let $T_j^{1} = \min\{T_{1j}, T_{2j}\}$ and $C_j$ be the time of censoring. An indicator variable for censoring is given by $\delta_j = I(T_j^{1} \leq C_j)$ = 1 (event) or 0 (censored). The observed event time for the individual is $T_j = \min\{T_{1j}, T_{2j}, C_j\}$. Let $X_j$ be the vector of covariates and $w_j$ be the survey weight of $j^{th}$ individual. Since only one event can happen at a given time, we observe $D_j = (T_j, C_j, \delta_j, T_{0j}, X_j, w_j)$. The weighted likelihood function takes the form: 

\begin{equation}\label{likelihood}
    L_w(\theta|D) = \prod_{j=1}^{N} \{ \lambda_1(t_j|\theta,X_j)^{\delta_j} \frac{S_1(t_j|\theta,X_j)}{S_1(t_{0j}|\theta,X_j)}\}^{w_j},   
\end{equation}

\noindent where $D = \{D_j,\;j = 1,...,N\}$, $\theta$ is the vector of all unknown parameters, and $\lambda_1$ and $S_1$ are as defined in \eqref{cudhazard} and \eqref{cudsurv}. 

\subsection{Prior Distributions}
The following independent prior distributions are assigned to the model parameters:
\[ \pi(\beta_0) = \text{N}(0, 100), \]
\[ \pi(\beta_i \mid \tau) = \text{double-exponential}\left(0, \frac{1}{\tau}\right),\: i > 0; \: \tau \sim \chi_1^2, \]
\[ \pi(\gamma) =  \pi(\gamma_1, \gamma_2, ..., \gamma_L) = \text{Dirichlet}(1,1,...,1). \]

\noindent A normal prior with a variance of 100 is assigned to $\beta_0$ to make it non-informative, the lasso prior is employed for the slope coefficients $\beta_i$ \cite{Van2019}, and the M-spline coefficients $\gamma_l$ in the baseline hazard function are jointly assumed to follow a non-informative Dirichlet distribution with concentration parameters set to 1 \cite{DirichletPrior1994}. 

\subsection{Posterior Distribution}
The joint posterior distribution is given by:
\begin{equation*}
    \pi(\theta|D) \propto \prod_{j=1}^{N} \{ \lambda_1(t_j|\theta,X_j)^{\delta_j} \frac{S_1(t_j|\theta,X_j)}{S_1(t_{0j}|\theta,X_j)}\}^{w_j}  \pi(\beta_0) \prod_{i>0}\pi(\beta_i|\tau) \pi(\tau) \prod_{l}\pi(\gamma_l).
\end{equation*}

\noindent The above distribution is estimated using an MCMC algorithm.

\section{Details of the Prediction Framework}
Once the above model is fitted, it can be used for prediction of CUD risk in a certain time frame for a person using cannabis $j^*$ (who may or may not be in training data), who is currently CUD-free and has covariate vector $X_{j^*}$. The predicted absolute risk for this individual within the age interval $(a, b]$ is given by:

\[ \hat{r}_{j^*}(a,b|X_{j^*},D) = \int \hat{r}_{j^*}(a,b|X_{j^*},\theta) \pi(\theta|D) d\theta, \]

\noindent where

\begin{equation}\label{abspostmean}
    \hat{r}_{j^*}(a,b|X_{j^*},\theta) = \int_a^b \hat{\lambda}_1(t|X_{j^*},\theta) \left(\frac{\hat{S}_1(t|X_{j^*},\theta)\hat{S}_2(t)}{\hat{S}_1(a|X_{j^*},\theta)\hat{S}_2(a)}\right) dt.
\end{equation}

\noindent We approximate $\hat{r}_{j^*}(a,b|X_{j^*},\theta)$ as the sum of the integrand over integer ages in the time interval of interest assuming that the risk remains constant in each 1-year interval. At $l^\text{th}$ MCMC iteration, we compute \eqref{abspostmean} at $\theta^{(l)}$ and then finally average them to get an constant approximate ${r}_{j^*}(a,b|X_{j^*},D)$.

We use the statistical software R \cite{R} for all computations. The M-spline basis is evaluated following the methodology described in \citet{Ramsay1988}, which is implemented in the splines2 R package \cite{splines2-paper}. To ensure identifiability of the M-spline coefficients, they are constrained to form a simplex \cite{Ramsay1988}. The MCMC algorithm is implemented using the R interface to Stan \cite{rstan}.

\section{Details of the Model Building Steps}

\begin{enumerate}
    \item \textbf{Construction of predictors}\\
    From the Add Health study questionnaires, we constructed 45 potential risk factors for CUD following \citet{Rajapaksha2022}. Among these predictors, 18 were from a single wave (cross-sectional) while 27 were measured across multiple waves (longitudinal). The longitudinal predictors were included in the model by extracting random intercepts from their respective linear mixed effects models as described in Section S1. Additionally, to ensure that all predictor values were measured before the occurrence of CUD, only the values that were measured at time points occurring strictly before the time of CUD onset or censoring were considered. Due to the substantial amount of missingness in some of the predictors, we employed the following structured approach to identify the most influential risk factors while maximizing the sample size for fitting regression models.
    
    \item \textbf{Initial screening}\\
    Initially, a weighted univariate Cox proportional hazard (PH) model subject to right censoring and left truncation was fitted for each predictor. The predictors with p-values below 0.25 in their respective model were retained. This resulted in 28 candidate predictors --- 13 cross-sectional and 15 longitudinal (random intercept summaries).
    
    \item \textbf{Multivariate screening}\\
    Further, weighted multivariate Cox PH models with different subsets of the 28 risk factors were fitted. The goal was to identify the most influential risk factors while ensuring that at least 80\% of the initial cohort had complete data on the selected risk factors. Of the 28 predictors, we retained 25 that had p-values below 0.25 in at least one of the models. Then we fitted a multivariate Cox PH model with these 25 predictors. We found that the coefficients of 4 predictors flipped signs as compared to their corresponding univariate PH models. We compared the performance of models with and without these 4 predictors using the C-index, which is an extended metric of the area under the receiver operating characteristic (AUC) in survival analysis \cite{cindex2}. There was a very limited impact of predictors with inconsistent signs on the model’s C-index, with the largest difference being 0.01. Thus, these variables were excluded from further consideration. This process identified 21 significant risk factors to be associated with CUD (see Table \ref{tab:S1}).
    
    \item \textbf{Bayesian model fitting}\\
    We then proceeded by fitting the Bayesian model with the 21 risk factors identified from the previous step. Since the prior distributions of the regression coefficients are continuous, the Bayesian framework does not lead to automatic variable selection. Two methods were employed for variable selection: credible interval and scaled neighborhood \cite{VarSelect2010}. The first method uses credible intervals for regression coefficients, selecting predictors whose intervals exclude 0 at a given confidence level (e.g., 70\%, 80\%, 95\%). The second method excludes predictors if the posterior probability that their regression coefficients lie within $\pm 1$ posterior standard deviation of 0 exceeds a certain threshold (e.g., 0.1, 0.6). Higher confidence levels and lower thresholds lead to stricter inclusion criteria, resulting in fewer predictors being selected.
    
    \item \textbf{Comparison of competing models}\\
    To determine the optimal model from the competing models, 5-fold cross-validation was conducted. We assessed each model's performance in two ways: measuring its discrimination using the AUC and evaluating its calibration by comparing expected and observed number of CUD events (E/O) \cite{Pfeiffer2017}. 

\end{enumerate}

\section{Recalibration of the Model on CHDS Data}

The lifetime prevalence of CUD in CHDS validation data is 16.48\%, which is notably higher than the 8.18\% prevalence in Add Health training data. Given this discrepancy and following \citet{Lakishka2023}, we performed model recalibration as suggested in literature \cite{Pfeiffer2017, recalibration1, recalibration2}. Specifically, we apply the logistic re-calibration to update
the model using the following steps:

\begin{enumerate}
    \item Estimate the absolute risks $r=(r_1,r_2,\ldots,r_n)$ for individuals in the CHDS data using the fitted Model 1.
    \item Calculate the logit of the risks from the previous step, denote them as $lr=(lr_1,lr_2,\ldots,lr_n)$, where 
    $lr_i=\log\left(\frac{r_i}{1-r_i}\right)$.
    \item Fit a logistic regression model on CHDS data with CUD status as the response and $lr$ as predictor. Obtain the estimated intercept of the model.
    \item Use the estimated intercept to update the logit of the risks $lr$ as
    $lr_{\text{new}}=lr+\text{intercept}$.
    \item Update the absolute risks as 
    $r_{\text{new}}=\frac{\exp(lr_{\text{new}})}{1+\exp(lr_{\text{new}})}$.
    \item Use the updated absolute risks $r_{\text{new}}$ to evaluate the prediction performance of the model.
\end{enumerate}


\newpage

\begin{table}[htbp]
  \refstepcounter{supplementarytable}
  \caption{List of predictors that passed the multivariate screening (Step 3 in Section S5).}
  \label{tab:S1}
  \begin{center}
  \begin{tabular}{lp{8cm}} \hline
  \textbf{Predictor } & \textbf{Description} \\ \hline
   Biological sex & Female, Male. \\
Race & White, Black, Native American, Asian, Other. \\

Neuroticism scale & Measure of the general tendency to experience negative emotions including mood swings and irritability. Higher values indicate a greater tendency to experience these feelings. \\

Conscientiousness scale & Measure of forward planning, organization, and ability to carry out tasks such as getting chores done right away and staying on top of responsibilities. A higher value implies greater conscientiousness. \\

Openness scale & Measure of openness to new experiences and imaginativeness including enjoying creative or unconventional activities. A higher value implies more openness. \\

Anxiety & Number of times experienced anxiety symptoms such as feeling hot all over suddenly for no apparent reason.\\

Depression & Number of times experienced depression symptoms such as feeling unable to shake off sadness, even with support from family or friends.\\

Delinquency & Number of times involved in delinquent activities such as deliberately damaging property that does not belong to you.\\

ACE & Number of adverse childhood experiences that occurred before age 18. \\

High school behavior & Measure of school-related problems not experienced by the respondent. A higher value implies fewer problems. \\

Supportive environment & Measure of perception of being supported by parents, teachers, friends, etc. A higher value implies greater support. \\

Peer cannabis use & Number of best friends (out of 3) who use cannabis at least once a month. \\

Peer smoking & Number of best friends (out of 3) who smoke at least 1 cigarette a day. \\

Parental education & Measure of highest education level achieved by any of the parents. A higher value implies more educated parents. \\

TV hours & Number of hours watched television per week. \\

Past 12-days alcohol use& Number of days that participants drank alcohol in the past 12 days.\\

Family structure & Family Structure of participants at age 5.\\

Welfare status & Indicator of whether a participant or anyone in the household is a welfare recipient. \\

Substance availability & Indicator of whether any substances are available to the participants.\\

Ever think suicide & Indicator of whether the participants ever think suicide. \\

Ever smoked & Indicator of whether the participants ever smoked an entire cigarette. \\
\hline
\end{tabular}
\end{center}
\end{table}

\begin{table}[htbp]
 \refstepcounter{supplementarytable} 
  \caption{Competing models as obtained in Step 4 of Section S5.}
  \label{tab:S2}
  \begin{center}
  \begin{tabular}{lcccccc} \hline
\textbf{Predictors} & \textbf{M1} & \textbf{M2} & \textbf{M3} & \textbf{M4} & \textbf{M5} & \textbf{M6} \\
\hline
Conscientiousness scale & $\checkmark$ & $\checkmark$ & $\checkmark$ & $\checkmark$ & $\checkmark$ & $\checkmark$ \\
Openness scale & $\checkmark$ & $\checkmark$ & $\checkmark$ & $\checkmark$ & $\checkmark$ & $\checkmark$ \\
Neuroticism scale & $\checkmark$ & $\checkmark$ & $\checkmark$ & $\checkmark$ & $\checkmark$ & $\checkmark$ \\
Delinquency scale & $\checkmark$ & $\checkmark$ & $\checkmark$ & $\checkmark$ & $\checkmark$ & $\checkmark$ \\
Biological sex &  & $\checkmark$ & $\checkmark$ & $\checkmark$ & $\checkmark$ & $\checkmark$ \\
Race & & & $\checkmark$ & $\checkmark$ & $\checkmark$ & $\checkmark$ \\
Peer cannabis use & & &  & $\checkmark$ & $\checkmark$ & $\checkmark$ \\
Parental education & & & & & $\checkmark$ & $\checkmark$ \\
Family structure & & & & & & $\checkmark$ \\
TV hours & & & & & & $\checkmark$ \\
\hline
\end{tabular}
\end{center}
\end{table}

\begin{table}[ht]
\refstepcounter{supplementarytable} 
  \caption{Prediction performance using 5-fold CV.}
  \label{tab:S3}
  \begin{center}
  \begin{tabular}{lllp{0.1cm}ll} \hline
  \textbf{Prediction} & \multicolumn{2}{c}{\textbf{Model 1}} &  & \multicolumn{2}{c}{\textbf{Model 2}} \\ 
  \cline{2-3} \cline{5-6}
  \textbf{age interval} & \textbf{AUC} & \textbf{E/O} &  & \textbf{AUC} & \textbf{E/O} \\
   \hline
   1-year after first cannabis use & 0.72  & 0.91  & &  0.72 & 0.92 \\
2-year after first cannabis use & 0.68  & 0.87  & &0.68  &  0.85  \\
3-year after first cannabis use & 0.69  & 0.87  & &0.70  &0.86    \\
5-year after first cannabis use & 0.68  & 0.95  &  &0.68 & 0.94   \\
8-year after first cannabis use & 0.68  & 1.00   & & 0.68 & 1.01     \\
11-year after first cannabis use & 0.68 & 0.93  && 0.68  &  0.94  \\
14-year after first cannabis use & 0.69 & 0.86  & &0.69   &0.85    \\
16-year after first cannabis use & 0.69 & 0.83  & &0.69  & 0.82   \\
20-year after first cannabis use & 0.69 & 0.82   & &0.69  &0.82   \\
(15, 16]            & 0.67 & 0.83  & &0.65  & 1.01   \\
(16, 17]               & 0.65 & 1.03   & & 0.66 &1.06  \\
(17, 18]              & 0.68 & 0.93   & &0.68  & 0.82  \\
(18, 19]               & 0.72 & 0.78   & &0.72  & 0.82  \\
(19, 20]              & 0.62 & 0.99  & &0.65 & 0.94   \\
(20, 21]               & 0.67 & 1.21  & &0.69  &1.24    \\
(15, 17]               & 0.66 & 0.90    & &0.65  & 0.97  \\
(16, 18]               & 0.66 & 1.01    &  &0.66 & 0.95 \\
(17, 19]               & 0.71 & 0.90    && 0.71  & 0.88  \\
(18, 20]               & 0.67 & 0.87  & &0.69  & 0.89   \\
(19, 21]              & 0.65 & 1.04   &  &0.67 & 1.05  \\
(20, 22]               & 0.67 & 1.32  & &0.68  & 1.39   \\
(15, 18]               & 0.66 & 0.92  & &0.65  & 0.94   \\
(16, 19]               & 0.69 & 0.99   && 0.68  & 0.98  \\
(17, 20]               & 0.69 & 0.95   & &0.69  & 0.94  \\
(18, 21]               & 0.67 & 0.94  && 0.68  & 0.98   \\
(19, 22]               & 0.65 & 1.16  & &0.67  & 1.16   \\
(20, 23]               & 0.65 & 1.39   && 0.65  & 1.44  \\
(15, 20]               & 0.68 & 0.94   && 0.67  & 0.98  \\
(16, 21]               & 0.67 & 1.03  && 0.67  & 1.06   \\
(17, 22]               & 0.68 & 1.06  && 0.68  &  1.07  \\
(18, 23]               & 0.66 & 1.09  & &0.67  & 1.14   \\
(19, 24]              & 0.63 & 1.23  & &0.65  & 1.25   \\
(20, 25]               & 0.65 & 1.23   & &0.65  & 1.26  \\
\hline
\end{tabular}
\end{center}
\end{table}

\begin{table}[htbp]
 \refstepcounter{supplementarytable} 
  \caption{Validation on Independent Add Health test data: Ratio of expected (E) to observed (O) number of CUD cases in 5-year prediction made at the age of first cannabis use.}
  \label{tab:S4}
  \begin{center}
  \begin{tabular}{lllllp{0.1cm}llll} \hline
   \textbf{Risk} & \multicolumn{4}{c}{\textbf{Model 1}} & & \multicolumn{4}{c}{\textbf{Model 2}} \\ 
  \cline{2-5} \cline{7-10}
   \textbf{groups} & \textbf{n} &
   \textbf{E} &\textbf{O} &\textbf{E/O}  & & \textbf{n} &
   \textbf{E} &\textbf{O} &\textbf{E/O} \\ \hline
Quartile 1	& 129 & 2.62 & 4 & 0.66 & & 128 & 2.26 & 3 & 0.75 \\
Quartile 2	& 130 & 4.04 & 3 & 1.35 & &  128 & 3.45 & 2 & 1.73 \\
Quartile 3	& 130 & 5.33 & 5 & 1.07 & &  128 & 4.60 & 5 & 0.92 \\
Quartile 4	& 130 & 9.48 & 10 & 0.95 & &  129 & 8.21 & 10 & 0.82\\
Biological sex \\
\hspace*{0.05in}Male	& 275 & 12.88 & 12 & 1.07 & &  271 & 10.98 & 11 & 1.00 \\
\hspace*{0.05in}Female	& 245 & 8.59 & 10 & 0.86 &  & 243 & 7.54 & 9 & 0.84\\
Conscientiousness scale 				\\
\hspace*{0.05in}Below median	& 305 & 13.96 & 13 & 1.07  & &  301 & 12.00 & 11 & 1.09 \\
\hspace*{0.05in}Above median	& 213 & 7.28 & 9 & 0.81  & & 211 & 6.32 & 9 & 0.70 \\
Neuroticism scale\\
\hspace*{0.05in}Below median	& 327 & 11.63 & 16 & 0.73 & &  323 & 9.93 & 15 & 0.66 \\
\hspace*{0.05in}Above median	& 188 & 9.71 & 6 & 1.62  & & 186 & 8.48 & 5 & 1.70\\
Openness scale				\\
\hspace*{0.05in}Below median	& 329 & 11.76 & 10 & 1.18  & & 325 & 10.01 & 8 & 1.25 \\
\hspace*{0.05in}Above median	& 189 & 9.66 & 12 & 0.80  & & 187 & 8.47 & 12 & 0.71\\ 
Delinquency scale				\\
\hspace*{0.05in}Below median	& 234 & 7.36 & 6 & 1.23  & & 233 & 6.46 & 5 & 1.29\\
\hspace*{0.05in}Above median	& 260 & 13.41 & 15 & 0.89  & & 257 & 11.51 & 15 & 0.77\\ 
Welfare				\\
\hspace*{0.05in}Recipient	 &  &  &  &  & & 171 & 6.37 & 8 & 0.80\\
\hspace*{0.05in}Nonrecipient &  &  &  &  & & 343 & 12.16 & 12 & 1.01\\ 
\hline
   \end{tabular}
  \end{center}
\end{table}

\begin{table}[htbp]
 \refstepcounter{supplementarytable} 
  \caption{Validation on independent Add Health test data: Ratio of expected (E) to observed (O) number of CUD cases in 5-year prediction made at age 16.}
  \label{tab:S5}
  \begin{center}
   \begin{tabular}{lllllp{0.1cm}llll} \hline
   \textbf{Risk} & \multicolumn{4}{c}{\textbf{Model 1}} & & \multicolumn{4}{c}{\textbf{Model 2}} \\ 
  \cline{2-5} \cline{7-10}
   \textbf{groups} & \textbf{n} &
   \textbf{E} &\textbf{O} &\textbf{E/O}  & & \textbf{n} &
   \textbf{E} &\textbf{O} &\textbf{E/O} \\ \hline
Quartile 1	& 94 & 2.44 & 3 & 0.81 & & 94 & 2.11 & 3 & 0.70\\
Quartile 2	& 95 & 3.39 & 3 & 1.13 & & 94 & 2.90 & 2 & 1.45 \\
Quartile 3	& 95 & 4.44 & 6 & 0.74 & & 94 & 3.82 & 7 & 0.55  \\
Quartile 4	& 95 & 8.76 & 8 & 1.09 && 95 & 7.65 & 8 & 0.96\\
Biological sex \\
\hspace*{0.05in}Male	& 203 & 11.75 & 13 & 0.90 & & 202 & 10.09 & 13 & 0.78 \\
\hspace*{0.05in}Female	& 177 & 7.30 & 7 & 1.04 && 176 & 6.41 & 7 & 0.92 \\
Conscientiousness scale 				\\
\hspace*{0.05in}Below median	& 226 & 12.47 & 12 & 1.04 &  & 225 & 10.78 & 12 & 0.90\\
\hspace*{0.05in}Above median	& 152 & 6.32 & 8 & 0.79 && 151 & 5.50 & 8 & 0.69\\
Neuroticism scale\\
\hspace*{0.05in}Below median	& 230 & 9.50 & 11 & 0.86 && 229 & 8.16 & 11 & 0.74 \\
\hspace*{0.05in}Above median	& 146 & 9.43 & 9 & 1.05 & & 145 & 8.24 & 9 & 0.92\\\
Openness scale				\\
\hspace*{0.05in}Below median	& 190 & 8.16 & 7 & 1.17 && 189 & 6.96 & 7 & 0.99\\
\hspace*{0.05in}Above median	& 188 & 10.83 & 13 & 0.83 && 187 & 9.51 & 13 & 0.73 \\ 
Delinquency scale				\\
\hspace*{0.05in}Below median	& 175 & 6.25 & 9 & 0.69 &  & 174 & 5.42 & 9 & 0.60 \\
\hspace*{0.05in}Above median	 & 190 & 12.31 & 11 & 1.12 &  & 189 & 10.65 & 11 & 0.97\\ 
Welfare				\\
\hspace*{0.05in}Recipient &&&&&	& 129 & 6.22 & 7 & 0.89 \\
\hspace*{0.05in}Nonrecipient &&&&&	& 249 & 10.28 & 13 & 0.79\\ 
\hline
   \end{tabular}
  \end{center}
\end{table}

\begin{table}[htbp]
 \refstepcounter{supplementarytable} 
  \caption{External validation of Model 1 on CHDS data: Ratio of expected (E) to observed (O) number of CUD cases in 5-year prediction made at the age of first cannabis use and at age 16.}
  \label{tab:S6}
  \begin{center}
  \begin{tabular}{lp{1.5cm}p{1.5cm}p{1.5cm}p{1.5cm}p{0.1cm}llll} \hline
   \textbf{Risk} & \multicolumn{4}{c}{\textbf{Prediction at the age of first cannabis use}} & & \multicolumn{4}{c}{\textbf{Prediction at age 16}} \\ 
  \cline{2-5} \cline{7-10}
   \textbf{groups} & \textbf{n} &
   \textbf{E} &\textbf{O} &\textbf{E/O}  & & \textbf{n} &
   \textbf{E} &\textbf{O} &\textbf{E/O} \\
   \hline
Quartile 1	& 159 & 8.96 & 2 & 4.48  && 42 & 4.79 & 4 & 1.20\\
Quartile 2	& 159 & 14.95 & 12 & 1.25 && 43 & 6.94 & 6 & 1.16\\
Quartile 3	& 159 & 20.88 & 22 & 0.95  && 42 & 8.63 & 12 & 0.72\\
Quartile 4	& 159 & 32.97 & 42 & 0.78&& 43 & 12.59 & 11 & 1.14 \\
Biological sex \\
\hspace*{0.05in}Male	& 319 & 42.96 & 57 & 0.75 & & 76 & 16.18 & 22 & 0.74	\\
\hspace*{0.05in}Female	& 318 & 34.83 & 21 & 1.66&& 95 & 16.84 & 11 & 1.53\\
Conscientiousness scale 				\\
\hspace*{0.05in}Below median	 & 410 & 56.60 & 55 & 1.03 & & 110 & 22.28 & 22 & 1.01 \\
\hspace*{0.05in}Above median	& 226 & 21.10 & 23 & 0.92&& 54 & 8.64 & 7 & 1.23 \\
Neuroticism scale\\
\hspace*{0.05in}Below median	& 241 & 26.02 & 25 & 1.04&& 77 & 13.75 & 16 & 0.86\\
\hspace*{0.05in}Above median	 & 300 & 43.07 & 46 & 0.94&& 74 & 15.93 & 13 & 1.23  \\
Openness scale				\\
\hspace*{0.05in}Below median	& 336 & 31.80 & 24 & 1.32&& 86 & 13.89 & 16 & 0.87\\
\hspace*{0.05in}Above median	& 299 & 45.87 & 53 & 0.87&& 83 & 18.91 & 17 & 1.11  \\ 
Delinquency scale				\\
\hspace*{0.05in}Below median	& 318 & 28.57 & 22 & 1.30& & 85 & 13.56 & 11 & 1.23\\
\hspace*{0.05in}Above median	 & 318 & 49.17 & 56 & 0.88&& 85 & 19.37 & 22 & 0.88\\ 
\hline
   \end{tabular}
  \end{center}
\end{table}

\begin{table}[htbp]
 \refstepcounter{supplementarytable} 
  \caption{External validation of Model 1 on CHDS data: AUC in predictions made at the age of first cannabis use and at specific ages.}
  \label{tab:S7}
  \begin{center}
  \begin{tabular}{p{4cm}p{1cm}cp{1cm}p{4cm}p{1cm}}
  \hline
  \multicolumn{2}{c}{\textbf{Prediction at the age of first cannabis use}} & & \multicolumn{3}{c}{\textbf{Prediction at specific ages}} \\ 
  \cline{1-2} \cline{4-6}
  \textbf{Follow-up} & \textbf{} & & \textbf{} & \textbf{Follow-up} & \textbf{} \\
   \textbf{duration} & \textbf{AUC} & & \textbf{Age} & \textbf{duration} & \textbf{AUC} \\
  \hline
   1 \text{Year} & 0.71 & & 16 & 1 \text{Year} & 0.69 \\
   2 \text{Year} & 0.71 & & 16 & 3 \text{Year} & 0.66 \\
   3 \text{Year} & 0.73 & & 16 & 5 \text{Year} & 0.65 \\
   5 \text{Year} & 0.75 & & 17 & 1 \text{Year} & 0.77 \\
   7 \text{Year} & 0.73 & & 17 & 3 \text{Year} & 0.65 \\
   8 \text{Year} & 0.73 & & 17 & 5 \text{Year} & 0.68 \\

   10 \text{Year} & 0.74 & & 18 & 1 \text{Year} & 0.61 \\
   11 \text{Year} & 0.74 & & 18 & 3 \text{Year} & 0.62 \\
   12 \text{Year} & 0.74 & & 18 & 5 \text{Year} & 0.66 \\
   14 \text{Year} & 0.74 & & 19 & 1 \text{Year} & 0.78 \\
   15 \text{Year} & 0.74 & & 19 & 3 \text{Year} & 0.73 \\
   16 \text{Year} & 0.74 & & 19 & 5 \text{Year} & 0.72 \\
\hline
   \end{tabular}
  \end{center}
\end{table}

\FloatBarrier
\begin{figure}[htbp]
    \centering
    \refstepcounter{supplementaryfigure}
    \includegraphics[scale=0.5]{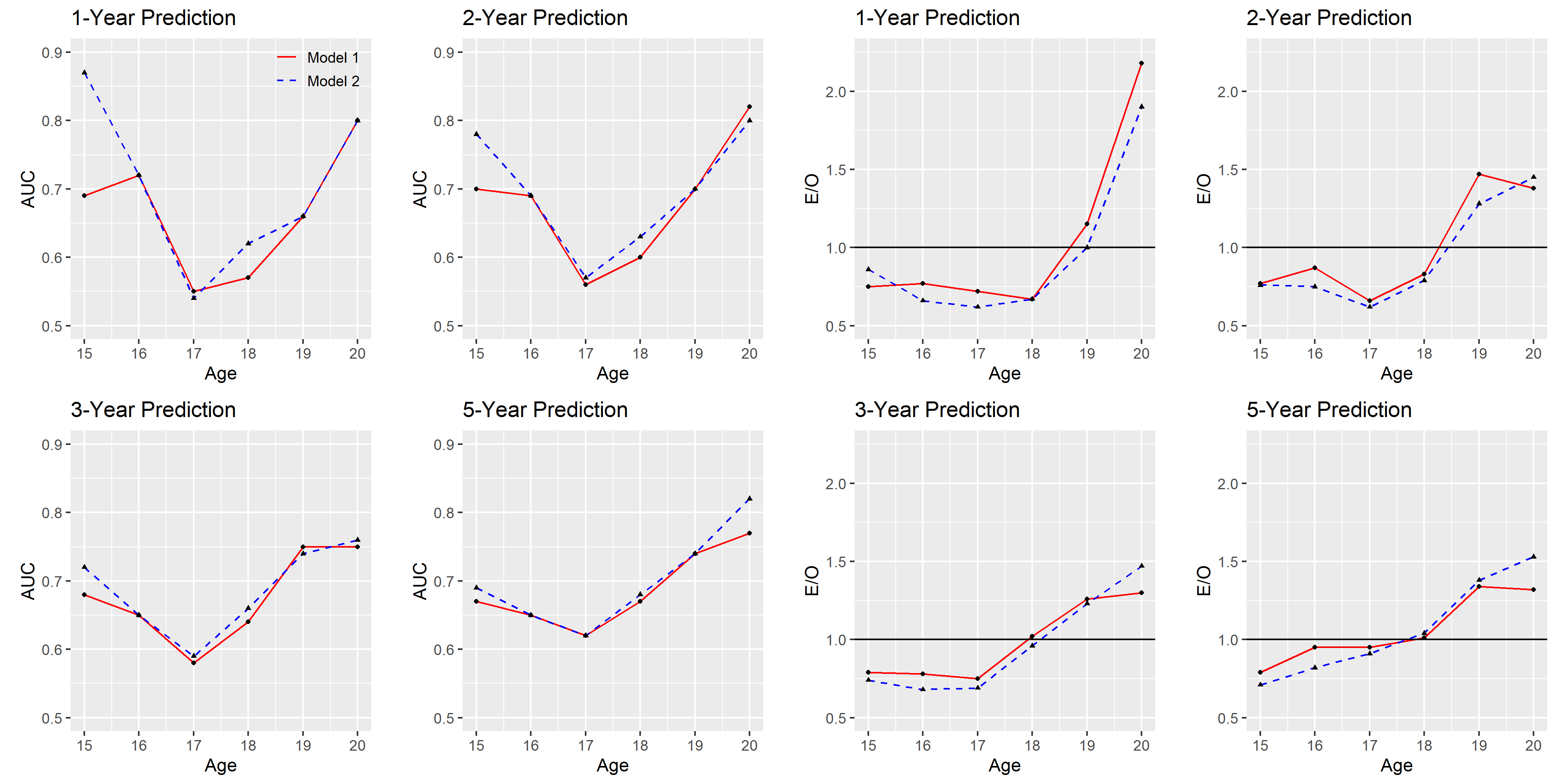}
    \caption{Validation on Add Health test data: AUC and E/O trajectories for CUD risk prediction within a certain number of years made at specific ages.}
    \label{fig:S1}
    
    \vspace{1cm} 
    
    \refstepcounter{supplementaryfigure}
    \includegraphics[scale=0.5]{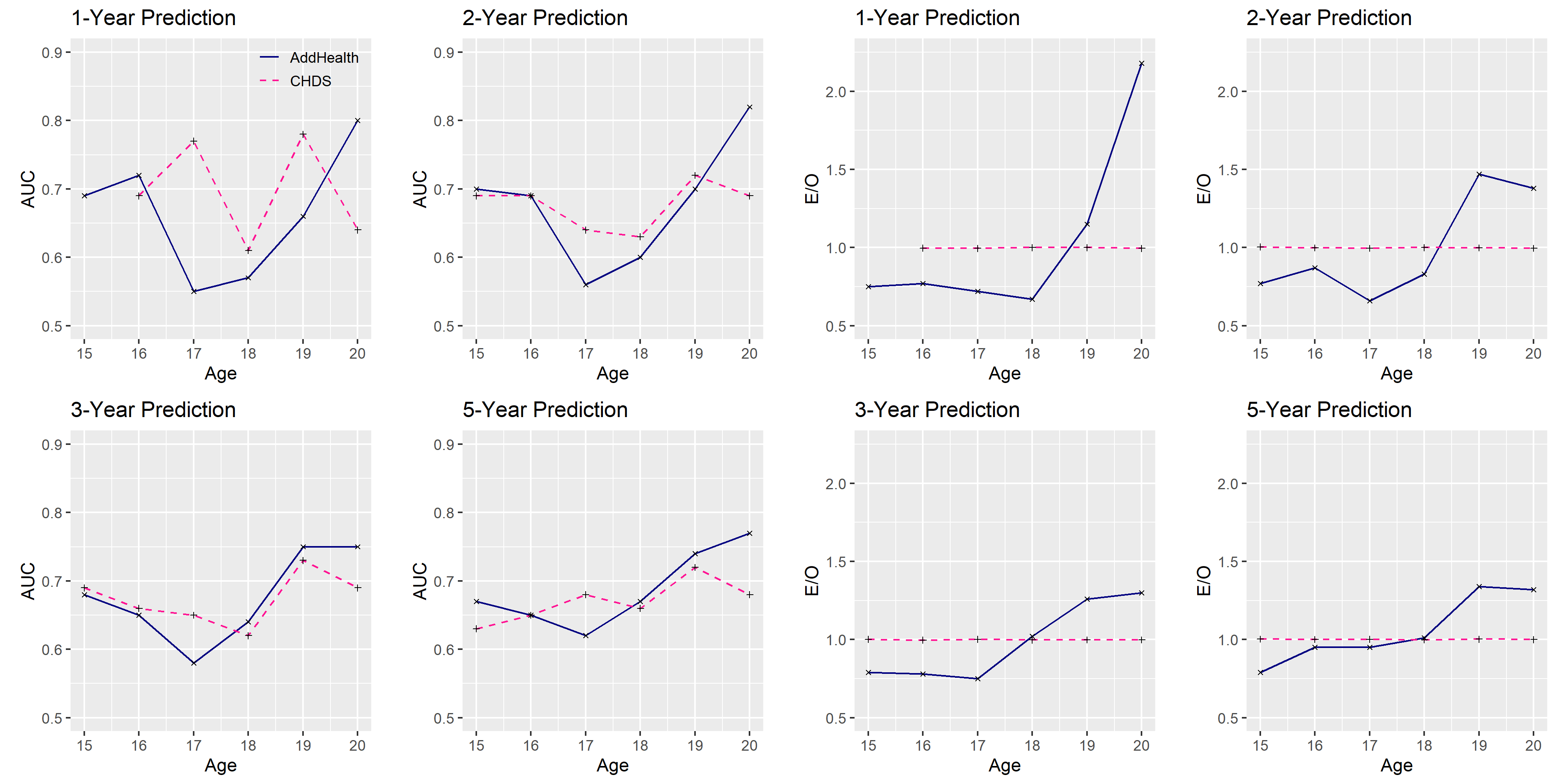}
    \caption{AUC and E/O trajectories of Model 1 for CUD risk prediction within a certain number of years made at specific ages for both validation datasets.}
    \label{fig:S2}
\end{figure}
\FloatBarrier

\bibliographystyle{unsrtnat}
\bibliography{ref}